\documentclass[sigconf]{acmart}
\DeclareUnicodeCharacter{FF09}{)}
\usepackage[T1]{fontenc}    
\usepackage{url}            
\usepackage{amsfonts}       
\usepackage{nicefrac}       
\usepackage{microtype}      
\usepackage{lipsum}
\usepackage{booktabs}
\usepackage{multirow}
\usepackage{rotating}
\usepackage{tabularx}
\usepackage{graphicx}       
\usepackage[utf8]{inputenc}
\usepackage{array}
\usepackage{caption}
\usepackage{adjustbox}
\usepackage{subcaption}
\usepackage{hyperref}
\usepackage{siunitx}
\usepackage{placeins}
\usepackage{float}
\newcolumntype{Y}{>{\centering\arraybackslash}X}
\graphicspath{{figs/}{media/}} 

\AtBeginDocument{%
  }

\setcopyright{none}
\copyrightyear{}
\acmYear{}
\acmDOI{}
\acmConference{}{}{}
\acmBooktitle{}
\acmISBN{}
\acmPrice{}
\begin{document}

\title{TCDE: Topic-Centric Dual Expansion of Queries and Documents with Large Language Models for Information Retrieval}
\renewcommand{\shorttitle}{TCDE: Topic-Centric Dual Expansion with LLMs}
\author{Yu Yang}
\affiliation{%
  \institution{Xi'an Jiaotong University}
  \city{Xi'an}
  \state{Shaanxi}
  \country{China}
}
\email{yyang@stu.xjtu.edu.cn}

\author{Feng Tian\textsuperscript{*}}
\thanks{\textsuperscript{*}Corresponding author}
\affiliation{%
  \institution{Xi'an Jiaotong University}
  \city{Xi'an}
  \state{Shaanxi}
  \country{China}
}
\email{fengtian@mail.xjtu.edu.cn}

\author{Ping Chen}
\affiliation{%
  \institution{University of Massachusetts Boston}
  \city{Boston}
  \state{MA}
  \country{United States}
}
\email{ping.chen@umb.edu}

\renewcommand{\shortauthors}{Yang et al.}

\begin{abstract}
Query Expansion (QE) enriches queries and Document Expansion (DE) enriches documents, and these two techniques are often applied separately. However, such separate application may lead to semantic misalignment between the expanded queries (or documents) and their relevant documents (or queries). To address this serious issue, we propose TCDE, a dual expansion strategy that leverages large language models (LLMs) for topic-centric enrichment on both queries and documents. In TCDE, we design two distinct prompt templates for processing each query and document. On the query side, an LLM is guided to identify distinct sub-topics within each query and generate a focused pseudo-document for each sub-topic. On the document side, an LLM is guided to distill each document into a set of core topic sentences. The resulting outputs are used to expand the original query and document. This topic-centric dual expansion process establishes semantic bridges between queries and their relevant documents, enabling better alignment for downstream retrieval models. Experiments on two challenging benchmarks, TREC Deep Learning and BEIR, demonstrate that TCDE achieves substantial improvements over strong state-of-the-art expansion baselines. In particular, on dense retrieval tasks, it outperforms several state-of-the-art methods, with a relative improvement of 2.8\% in NDCG@10 on the SciFact dataset. Experimental results validate the effectiveness of our topic-centric and dual expansion strategy.
\end{abstract}




\maketitle

\section{Introduction}

In information retrieval (IR), query expansion (QE) and document expansion (DE) are effective strategies to mitigate the vocabulary mismatch problem ~\cite{belkin1982ask}. These techniques enhance the semantic representations of queries or documents by incorporating additional contextual information ~\cite{carpineto2012survey}. Traditional QE methods, such as pseudo-relevance feedback (PRF) ~\cite{robertson1990term,salton1990improving,jones2006generating,lavrenko2017relevance,kuzi2016query}, expand queries using terms extracted from the top-ranked documents returned in the first-stage retrieval. Similarly, DE techniques expand documents by leveraging neighboring content ~\cite{billerbeck2005document,tao2006language,Efron2012ImprovingRO}, or by utilizing external collections ~\cite{10.1145/3077136.3080716}.

Recent advances in large language models (LLMs) ~\cite{brown2020language} have shown strong potential in expanding queries via diverse semantic content, including generated passages ~\cite{wang2023query2doc}, chain-of-thought reasoning ~\cite{jagerman2023query}, and multi-format augmentations such as keyword lists, named entities, factual statements, and summaries ~\cite{10.1145/3539618.3591992}. In parallel, document expansion has also been explored using LLMs. For instance, Nogueira et al. ~\cite{nogueira2019doc2query} generate queries using Seq2Seq models to enrich documents, while Ma et al. ~\cite{ma2023pre} leverage LLMs to generate relevant queries to expand documents.  However, due to their inherently asymmetric design, such expansion methods lead to semantic misalignment between expanded queries (or documents) and their relevant documents (or queries).

\begin{figure}[htbp]
    \centering
    \begin{subfigure}{\linewidth}
        \centering
        \includegraphics[trim=15 10 15 10, clip, width=\linewidth]{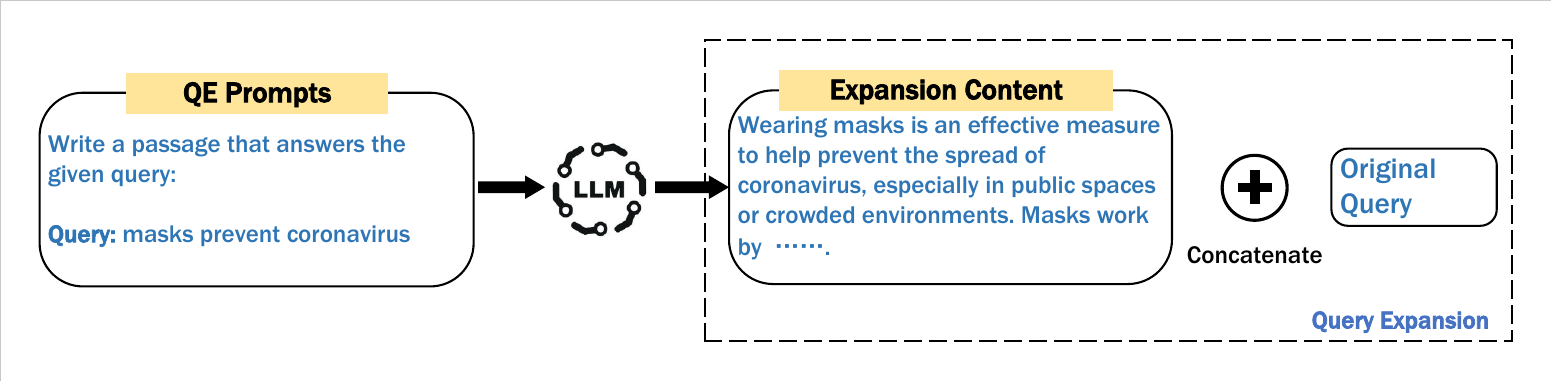}
        \caption{Query Expansion with LLM.}
        \label{fig:sub1}
    \end{subfigure}

    \vspace{0.5em}
    \begin{subfigure}{\linewidth}
        \centering
        \includegraphics[trim=15 10 15 10, clip, width=\linewidth]{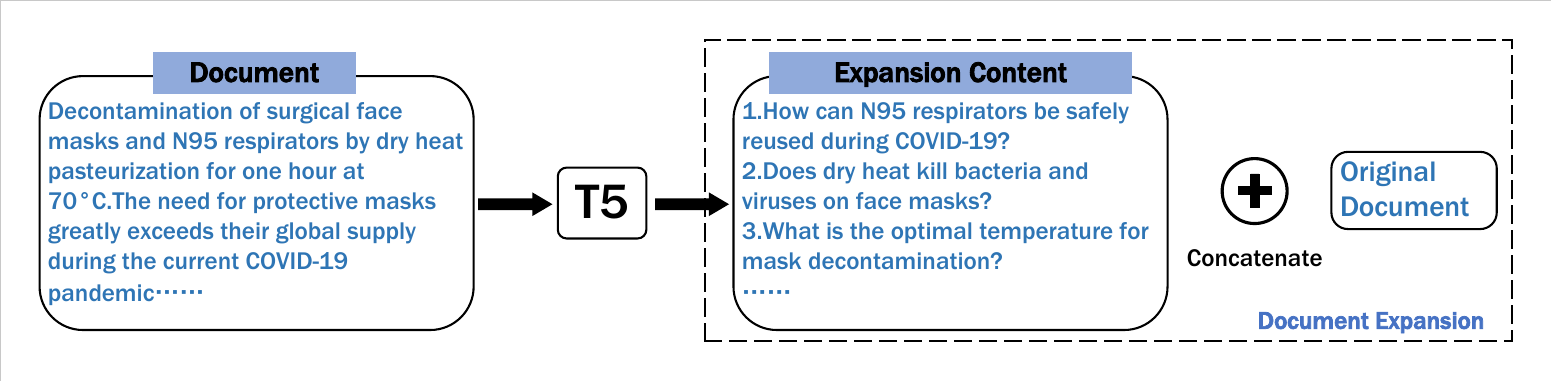}
        \caption{Document Expansion with a trained Seq2Seq model.}
        \label{fig:sub2}
    \end{subfigure}
    
    \vspace{0.5em}
    \begin{subfigure}{\linewidth}
        \centering
        \includegraphics[trim=15 10 15 10, clip, width=\linewidth]{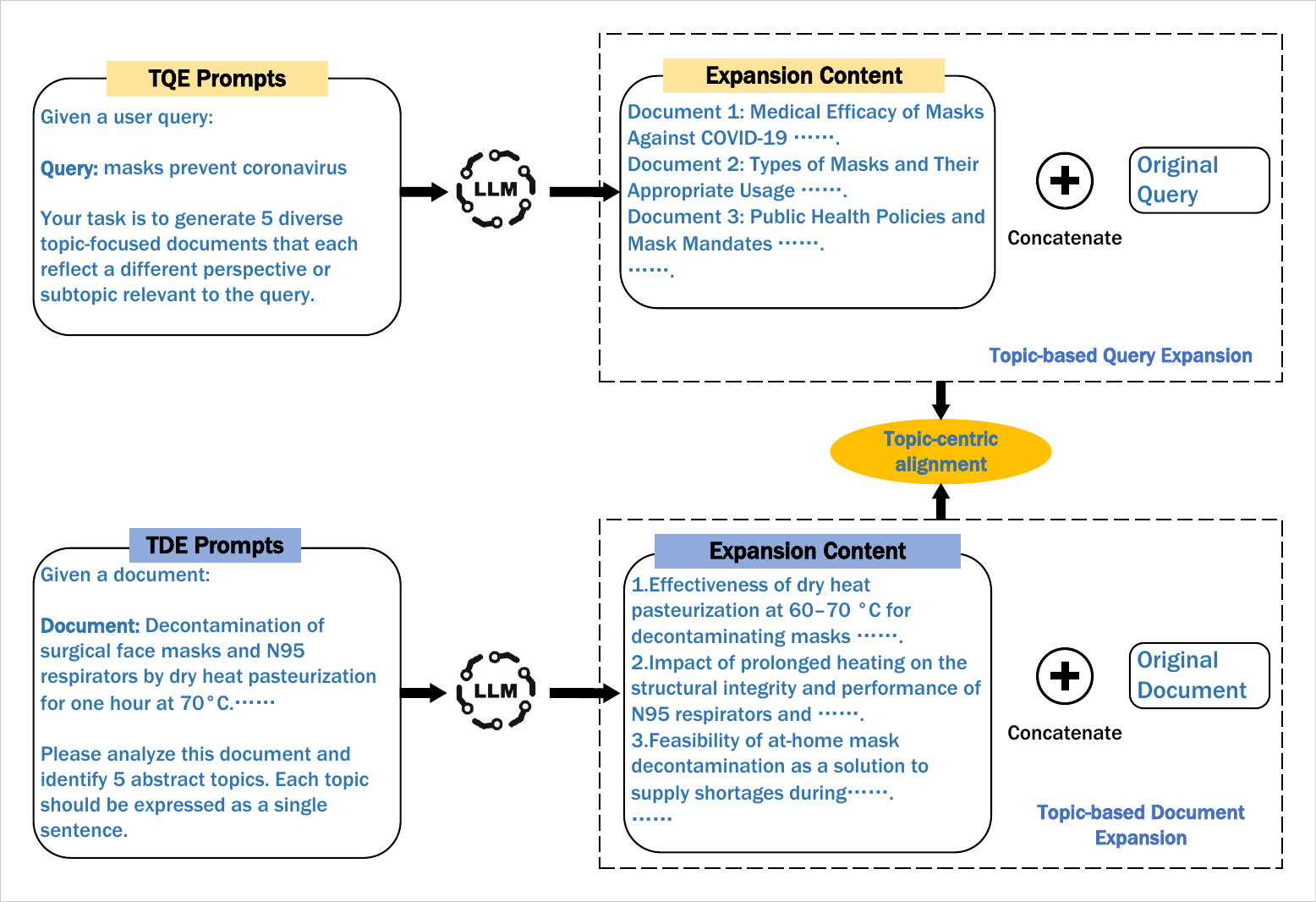}
        \Description{Overview of different expansion techniques in information retrieval.}
        \caption{Our topic-centric dual expansion.}
        \label{fig:sub3}
    \end{subfigure}

    \caption{Overview of different expansion techniques in information retrieval: 
(a) Query expansion using LLMs, 
(b) Document expansion using a Seq2Seq model, 
and (c) Our proposed topic-centric dual expansion that simultaneously enriches both queries and documents.}
    \label{fig:Expansion_illustration}
\end{figure}

To address this semantic misalignment problem, we propose TCDE, a novel training-free framework that leverages LLMs to perform topic-centric, dual expansion of both queries and documents. As illustrated in \autoref{fig:Expansion_illustration}, \autoref{fig:sub1} and \autoref{fig:sub2} show conventional single-sided expansion approaches, query expansion using an LLM and document expansion using a trained Seq2Seq model, respectively. In contrast, \autoref{fig:sub3} presents our proposed method, which synergistically expands both queries and documents through topic-centric expansion. Since TCDE performs topic-centric dual expansion, and topics are inherently more abstract and concise than typical pseudo-documents, this design facilitates more effective semantic alignment between queries and documents, thereby enhancing downstream retrieval performance.



\begin{figure*}[htbp]
    \centering
    \includegraphics[trim=3pt 8pt 12pt 8pt, clip, width=\textwidth]{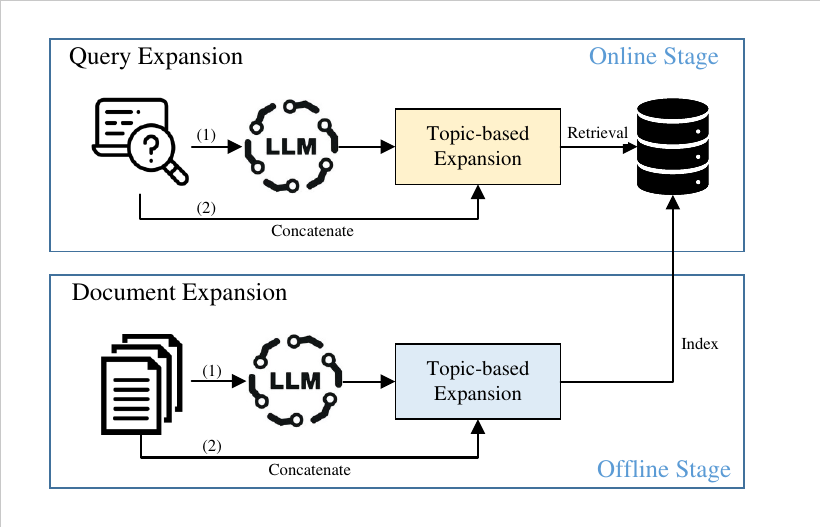}
    \Description{Overview of the TCDE framework.}
    \caption{Overview of the TCDE framework. Online stage: each incoming query $q$ is expanded by the LLM into $N$ topic-focused pseudo-documents (TQE) and concatenated with the original query to form $q^+$. Offline stage: each document $d$ is summarized by the LLM into $N$ concise topic sentences (TDE) and concatenated with the original document to obtain $d^+$, which is then indexed. Retrieval is performed by matching $q^+$ against the index of $d^+$, enforcing topic-centric alignment between queries and documents.
}
    \label{fig2:framework}
\end{figure*}

In summary, our key contributions are as follows.

\begin{itemize}
    \item We propose TCDE, a training-free framework that leverages large language models to synergistically expand both queries and documents. By performing dual expansion, TCDE effectively mitigates semantic misalignment in traditional asymmetric approaches that expand only queries or documents.
    
    \item We introduce a novel topic-centric expansion strategy that better captures semantic connection between queries and documents, outperforming traditional pseudo-document-based expansion approaches.
    
    \item Extensive experiments on the BEIR ~\cite{thakur2021beir} and TREC DL ~\cite{craswell2022overview} benchmarks demonstrate that TCDE consistently improves sparse and dense retrieval performance in diverse datasets.
\end{itemize}

\section{Related work}

\subsection{Information Retrieval}
Information retrieval (IR) aims to satisfy a user’s information need by ranking a large collection of documents $
    \mathcal{C} = \left\{d_{1}, d_{2},...,d_{|\mathcal{C}|}\right\}$ according to their relevance to a given query $q$. A retrieval system learns a scoring function $S(q,d)$ that assigns a relevance score to each query–document pair $(q,d)$, where $d \in \mathcal{C}$, and ranks documents in descending order according to these scores.

Information retrieval mainly follows two paradigms: sparse retrieval and dense retrieval.
In sparse retrieval, queries and documents are represented as high-dimensional sparse vectors, typically under a Bag-of-Words (BoW) model, where each dimension corresponds to a unique vocabulary term. An example is BM25 \cite{robertson2009probabilistic}, which is efficient and effective but suffers from the vocabulary mismatch problem, as it cannot capture semantic relationships beyond exact term matches.

Dense retrieval addresses this limitation by encoding queries and documents into low-dimensional dense vectors (embeddings) via deep neural networks \cite{karpukhin2020dense, xiong2021approximate}. Relevance is computed as the similarity between these embeddings, enabling semantic matching that bridges the vocabulary gap.

\subsection{Query Expansion}
Query expansion (QE) is a classic technique for enriching original queries. Early approaches centered on pseudo-relevance feedback (PRF) \cite{salton1990improving,10.1145/502585.502654}, which extracts informative terms from top-ranked documents to expand original queries in the initial retrieval stage. With the rise of neural representations, researchers began leveraging pre-trained word embeddings to capture semantic relationships. A representative example is the work of Kuzi et al. \cite{kuzi2016query}, which identifies semantically similar terms in vector space to serve as expansion terms, marking a shift from purely lexical to semantic-level query expansion. 
The emergence of Large Language Models (LLMs) has led to another paradigm shift. Due to their extensive pre-training and strong instruction-following ability, LLMs can generate fluent and diverse expansion content that substantially enriches the semantics of the original query. For example, HyDE \cite{gao-etal-2023-precise} proposed the generation of a hypothetical response document as a form of expansion. This idea was extended by Query2doc \cite{wang2023query2doc}, which directly used LLM-generated documents as expansion text, demonstrating further gains in retrieval effectiveness. Some works go beyond answer-based generation. For example, chain-of-thought (CoT) reasoning has been adopted for content \cite{jagerman2023query}. Likewise, GRF \cite{10.1145/3539618.3591992} prompts LLMs to produce a multifaceted expansion that includes keywords, entities, and related questions. To mitigate potential hallucinations in LLM output, CSQE \cite{lei-etal-2024-corpus} introduces a retrieval-augmented strategy, where an LLM expand queries using content drawn from the initial documents retrieved.

\subsection{Document Expansion}
In contrast to QE, Document Expansion (DE) enriches the documents themselves to better match potential user queries. Early work by Billerbeck et al. \cite{billerbeck2005document} expanded documents using terms from the local context or external collections. Tao et al. \cite{tao-etal-2006-language} proposed language model-based approaches to infer related terms. A landmark contribution, Doc2Query \cite{DBLP:journals/corr/abs-1904-08375}, used a Seq2Seq model to generate potential queries for each document and append them to the original document, significantly improving sparse retrieval performance. This idea was further enhanced by docT5query \cite{nogueira2019doc2query}, which employed the more powerful T5 model to generate higher quality expansions and examined the effect of varying the number of generated queries in a follow-up study \cite{10.1007/978-3-031-28238-6_31}. Recently, LLM-based DE has emerged, with Ma et al. \cite{ma2023pre} and Ye et al. \cite{ye2024enhancing} leveraging large language models to synthesize rich, contextually relevant expansions, demonstrating the promising potential of LLMs for document expansion.


\section{Method}
In this section, we present our proposed Topic-Centric Dual Expansion (TCDE). Unlike existing studies that typically apply LLMs to either QE or DE independently, TCDE is a framework that unifies QE and DE in a dual and symmetric expansion strategy. The overall architecture of TCDE is illustrated in \autoref{fig2:framework}. Our method consists of two core synergistic components: Topic-Centric Query Expansion and Topic-Centric Document Expansion, both powered by a Large Language Model (LLM).

\subsection{Topic-Centric Query Expansion (TQE)}
A user query may contain multiple latent topics. To capture these topics and better satisfy the user’s information need, we employ a large language model (LLM). Specifically, for a given query $q$, we prompt the LLM to generate $N$ distinct, yet related, topic-centric expansion texts, with the prompt shown in \autoref{tab:tqe_prompt}. This process can be formulated as follows.
\begin{equation}
    D_{Topic}=\left\{d_{t1},d_{t2},\ldots,d_{tN}\right\}=LLM(q)
\end{equation}
where $D_{Topic}$ is a set of $N$ generated topic-centric pseudo-documents for query expansion and $N$ represents the total number of expansion topics. When constructing the final expanded query, it is crucial to balance the influence of the original query and the newly generated content. To preserve the original search intent and enhance its weight, we adopt the strategy proposed by Query2doc \cite{wang2023query2doc} and repeat the original query five times. Consequently, the expanded query $q^+$ is constructed by concatenating the repeated original query with the generated topic documents:
\begin{equation}
   q^+=\mathrm{concat}(q,q,q,q,q,D_{Topic}) 
\end{equation}

\begin{table}[htbp]
    \centering
    \begin{tabularx}{\linewidth}{X} 
    \toprule
      \textbf{TQE prompt} \\ 
    \midrule
      Given a user query: [Query] \\
      Your task is to generate $N$ diverse topic-focused documents that each reflect a different perspective or subtopic relevant to the query. \\
    \bottomrule
    \end{tabularx}
    \caption{The prompt for Topic-Centric Query Expansion (TQE).} 
    \label{tab:tqe_prompt}
\end{table}

\subsection{Topic-Centric Document Expansion (TDE)}
In contrast to query expansion, we take a cautious approach to document expansion. Since the source text is already informative, producing synthetic expansions risks topic drift. Instead, our DE module prompts an LLM to extract $N$ relevant topics from the original document $d$, each summarized in one concise sentence. The prompt is shown in \autoref{tab:tde_prompt}. The process is defined as follows.
\begin{equation}
    S_{Topic}=\left\{s_{t1},s_{t2},\ldots,s_{tN}\right\}=LLM(d)
\end{equation}
where $S_{Topic}$  is a set of $N$ generated topic sentences. The number of generated topics, $N$, is kept consistent with the query expansion stage to maintain symmetry.
Finally, the expanded document $d^+$ is formed by appending the generated topic sentences to the original document:
\begin{equation}
    d^+=\mathrm{concat}(d,S_{Topic})
\end{equation}

\begin{table}[htbp]
    \centering
    \begin{tabularx}{\linewidth}{X} 
    \toprule
      \textbf{TDE prompt} \\ 
    \midrule
      Given a document: [Document] \\
      Please analyze this document and identify 5 abstract topics. Each topic should be expressed as a single sentence. \\
    \bottomrule
    \end{tabularx}
    \caption{The prompt for Topic-Centric Document Expansion (TDE).} 
    \label{tab:tde_prompt}
\end{table}

Although the Topic-Centric Document Expansion only adds a few topic sentences, this design offers several advantages. First, it avoids topic drift in the representation of documents. Second, these topic sentences, which summarize the thematic content of the documents, are aligned with the topic-centric expansions of queries, thereby enabling more effective retrieval.

\subsection{Topic-centric alignment}

\begin{figure}[htbp]
    \centering
    \begin{subfigure}{\linewidth}
        \centering
        \includegraphics[trim=15 10 15 10, clip, width=\linewidth]{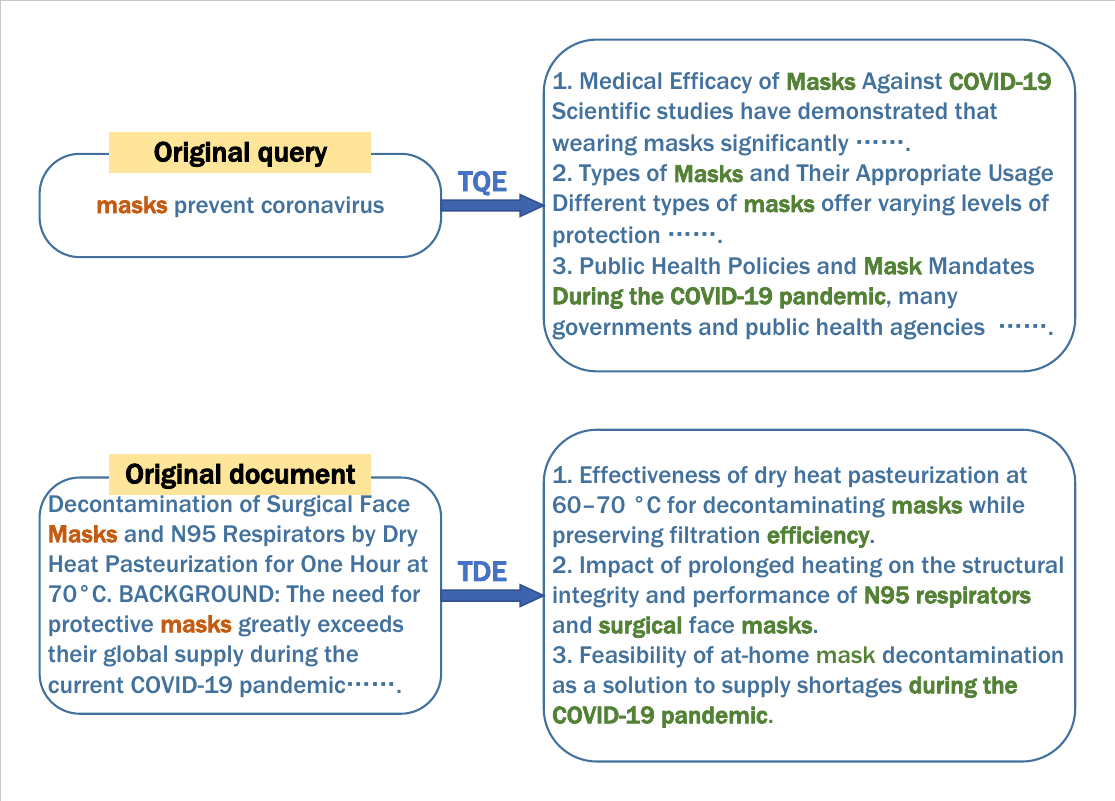}
        \caption{\textbf{Topic-centric alignment at the lexical level.} We highlight keywords overlaps to show lexical alignment: red bold marks overlaps between the original query and document, and green bold marks overlaps introduced by their topic-centric expansions.}
        \label{fig:lexical-level}
    \end{subfigure}

    \vspace{0.5em}
    \begin{subfigure}{\linewidth}
        \centering
        \includegraphics[width=\linewidth]{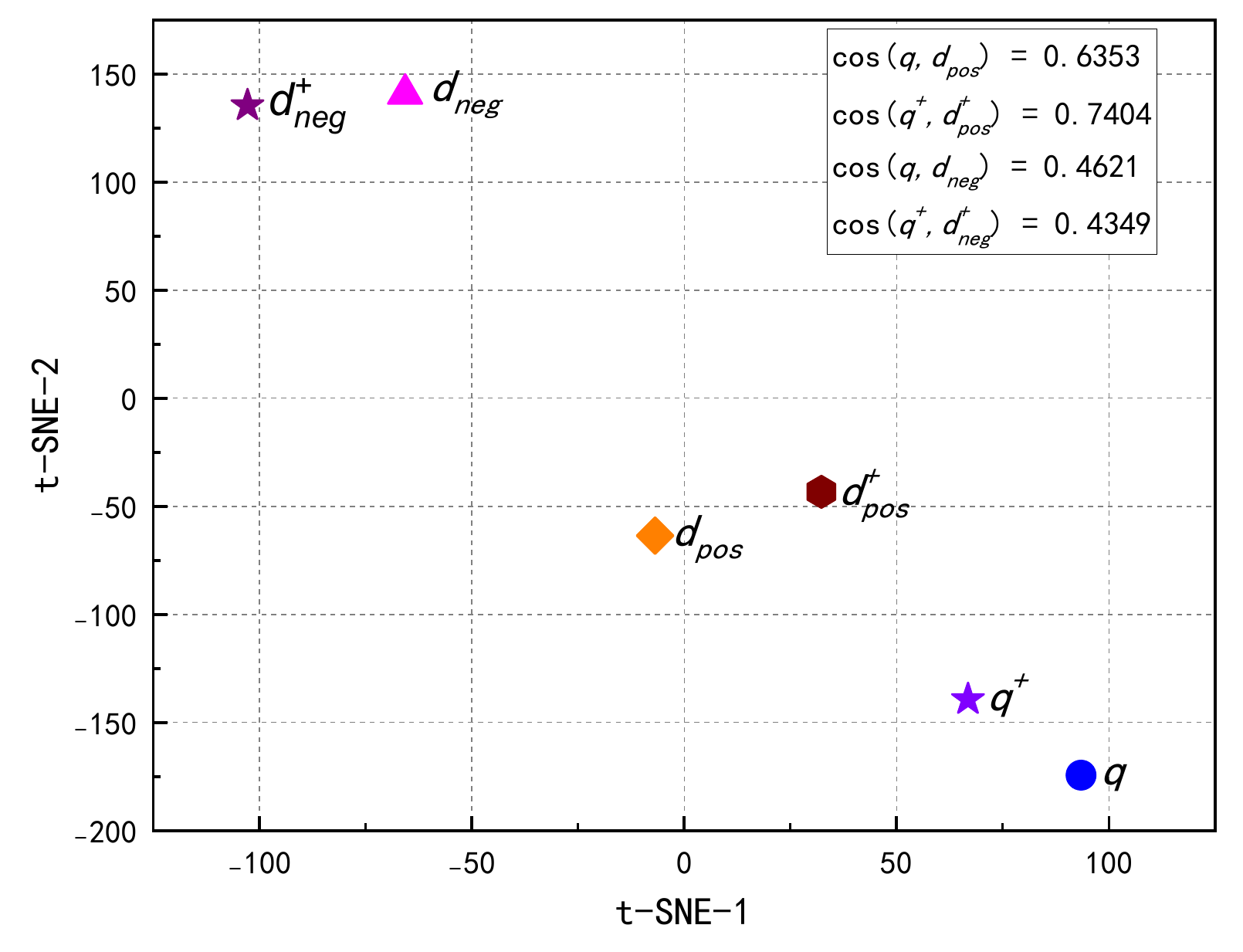}
        \caption{\textbf{Topic-centric alignment at the semantic level. t-SNE visualization of embeddings for the query and documents before
  \((q, d_{\mathrm{pos}}, d_{\mathrm{neg}})\) and after topic-centric expansion
  \((q^{+}, d_{\mathrm{pos}}^{+}, d_{\mathrm{neg}}^{+})\).
  The expansion pulls the positive pair closer in embedding space
  \(\cos(q, d_{\mathrm{pos}})=0.6353 \rightarrow \cos(q^{+}, d_{\mathrm{pos}}^{+})=0.7404\),
  while keeping the negative pair relatively distant
  \(\cos(q, d_{\mathrm{neg}})=0.4621\), \(\cos(q^{+}, d_{\mathrm{neg}}^{+})=0.4349\).
  This indicates improved semantic alignment that benefits dense retrieval.} }
        \label{fig:semantic-level}
    \end{subfigure}

    \caption{Topic-centric alignment at the lexical and semantic levels.}
    \label{fig:topic-centric alignment}
\end{figure}

\paragraph{\textbf{Topic-centric alignment at the lexical level.}}
As shown in \autoref{fig:lexical-level}, topic-centric query and document expansion increases the keyword overlap between the expanded query and document, thereby achieving lexical alignment and improving performance in sparse retrieval.

\paragraph{\textbf{Topic-centric alignment at the semantic level.}}
As shown in \autoref{fig:semantic-level}, our topic-centric dual expansion (TCDE) also achieves semantic alignment: it draws the positive query--document pair closer in embedding space while pushing the negative pair farther apart.
Formally, let $(q, d_{\mathrm{pos}}, d_{\mathrm{neg}})$ denote a query with a relevant and an irrelevant document, and let $S(\cdot,\cdot)$ be the model’s relevance scoring function (e.g., cosine similarity for dense retrieval). After applying TCDE, the triple becomes $(q^{+}, d^{+}_{\mathrm{pos}}, d^{+}_{\mathrm{neg}})$. Because TCDE pulls the positive pair closer and pushes the negative pair farther apart, we have
\begin{equation}
  S\!\big(q^{+}, d^{+}_{\mathrm{pos}}\big) \;>\; S(q, d_{\mathrm{pos}}) 
  \quad \text{and} \quad
  S\!\big(q^{+}, d^{+}_{\mathrm{neg}}\big) \;<\; S(q, d_{\mathrm{neg}}).
\end{equation}

\noindent
In summary, TCDE performs a dual, synergistic expansion by generating topic-focused pseudo-documents for queries (TQE) and concise topic summaries for documents (TDE). This topic-centric process strengthens both lexical and semantic alignment between queries and relevant documents, thereby improving the effectiveness of both sparse and dense retrieval. The next section empirically validates this approach.

\section{Experiments}
This section details our experimental setup for evaluating the effectiveness of TCDE in sparse and dense retrieval settings. We describe the datasets, models, and baseline methods used for comparison.

\subsection{Experiment Settings}

\textbf{Datasets and Metrics}. We evaluate on (1) three web-search datasets: MS MARCO ~\cite{nguyen2017ms}, TREC DL 2019 ~\cite{craswell2022overview}, and TREC DL 2020 ~\cite{craswell2022overview}; and (2) the zero-shot BEIR benchmark ~\cite{thakur2021beir}, which covers a variety of domains (e.g., scientific and biomedical). From BEIR, we use the following datasets: ArguAna, Climate-FEVER, DBPedia-Entity, FEVER, FiQA, HotpotQA, NFCorpus, NaturalQuestions, SCIDOCS, SciFact, TREC-COVID, and Webis-Touche2020. Detailed statistics are given in Table ~\ref{tab:dataset_stats}.

For evaluation metrics, we adopt the widely-used NDCG@K, MAP@K and Recall@K.

\begin{table}[htbp]
    \centering
    \begin{tabular}{l l r r} 
        \toprule
        \textbf{Benchmark} & \textbf{Dataset} & \textbf{Test Queries} & \textbf{Corpus} \\
        \midrule
        \multirow{3}{*}{TREC DL}
            & MS MARCO Dev        & 6,980 & 8,841,823 \\
            & TREC DL'19 Passage & 43    & 8,841,823 \\
            & TREC DL'20 Passage & 54    & 8,841,823 \\
        \midrule
        \multirow{12}{*}{BEIR}
            & arguana        & 1,406 & 8,674 \\
            & climate-fever  & 1,535 & 5,416,593 \\
            & dbpedia-entity & 400   & 4,635,922 \\
            & fever          & 6,666 & 5,416,568 \\
            & fiqa           & 648   & 57,638 \\
            & hotpotqa       & 7,405 & 5,233,329 \\
            & nfcorpus       & 323   & 3,633 \\
            & nq             & 3,452 & 2,681,468 \\
            & scidocs        & 1,000 & 25,657 \\
            & scifact        & 300   & 5,183 \\
            & TREC-COVID     & 50    & 171,332 \\
            & Webis-Touche2020          & 49    & 382,545 \\
        \bottomrule
    \end{tabular}
    \caption{Statistics of the datasets used for evaluation. Corpus size indicates the number of documents.}
    \label{tab:dataset_stats}
\end{table}

\begin{table*}[!htbp]
    \centering
    \setlength{\tabcolsep}{4pt} 
    \resizebox{0.9\textwidth}{!}{ 
    \begin{tabular}{l|rrr|rrr|rrr}
        \toprule
        \multirow{2}{*}{\textbf{Methods}} & \multicolumn{3}{c|}{\textbf{MS MARCO Dev}} & \multicolumn{3}{c|}{\textbf{TREC DL'19}} & \multicolumn{3}{c}{\textbf{TREC DL'20}} \\
         & N@10 & M@10 & R@1k & N@10 & M@10 & R@1k & N@10 & M@10 & R@1k \\
        \midrule
        \multicolumn{10}{c}{\textit{\textbf{Sparse Retrieval}}} \\
        \midrule
        BM25               & 0.2256 & 0.1762 & 0.8566 & 0.4734 & 0.1059 & 0.7322 & 0.4787 & 0.1362 & 0.7456 \\
        \quad + docT5query & \textbf{0.3123} & \textbf{0.2482} & \textbf{0.9316} & 0.5938 & 0.1319 & 0.7844 & 0.5902 & 0.1815 & 0.7808 \\
        \quad + Q2D        & 0.1679 & 0.1306 & 0.8462 & 0.5309 & 0.1134 & 0.7758 & 0.5300 & 0.1538 & 0.7459 \\
        \quad + Q2C        & 0.2057 & 0.1606 & 0.8793 & 0.5839 & 0.1229 & 0.7945 & 0.5224 & 0.1426 & 0.7621 \\
        \quad + GRF        & 0.2358 & 0.1849 & 0.9172 & 0.6620 & 0.1553 & \textbf{0.8739} & 0.6143 & 0.1931 & \textbf{0.8604} \\
        \quad + CSQE       & 0.1701 & 0.1330 & 0.8450 & 0.6597 & 0.1497 & 0.8551 & \textbf{0.6423} & \textbf{0.1994} & 0.8567 \\
        \quad + TCDE       & 0.2549 & 0.2001 & 0.9270 & \textbf{0.6657} & \textbf{0.1626} & 0.8637 & 0.6039 & 0.1829 & 0.8588 \\
        \midrule
        \multicolumn{10}{c}{\textit{\textbf{Dense Retrieval}}} \\
        \midrule
        E5-Base             & \textbf{0.4134} & \textbf{0.3439} & \textbf{0.9757} & 0.6943 & 0.1416 & 0.7735 & 0.6714 & 0.1974 & 0.7735 \\
        \quad + docT5query  & 0.4093 & 0.3403 & 0.9744 & 0.7025 & 0.1437 & 0.7612 & 0.6974 & 0.2095 & 0.7668 \\
        \quad + Q2D         & 0.2862 & 0.2303 & 0.9356 & 0.6743 & 0.1524 & 0.8048 & 0.5839 & 0.1734 & 0.7401 \\
        \quad + Q2C         & 0.2370 & 0.1880 & 0.8792 & 0.5494 & 0.1111 & 0.6816 & 0.4960 & 0.1390 & 0.6481 \\
        \quad + GRF         & 0.2091 & 0.1648 & 0.8521 & 0.5903 & 0.1170 & 0.7267 & 0.4973 & 0.1328 & 0.6538 \\
        \quad + CSQE        & 0.2906 & 0.2331 & 0.9416 & 0.6816 & 0.1497 & 0.8381 & 0.6539 & 0.1960 & 0.8127 \\
        \quad + TCDE        & 0.3857 & 0.3155 & 0.9742 & \textbf{0.7235} & \textbf{0.1630} & \textbf{0.8554} & \textbf{0.7178} & \textbf{0.2222} & \textbf{0.8268} \\
        \bottomrule
    \end{tabular}
    }
    \caption{Evaluation results on the MS MARCO passage retrieval benchmark, including MS MARCO Dev, TREC DL'19, and TREC DL'20. We report NDCG@10 (N@10), MAP@10 (M@10), and Recall@1000 (R@1k). Bold numbers indicate the best performance within each block (sparse or dense retrieval).}
    \label{tab:results_msmarco}
\end{table*}

\textbf{LLM for Expansion}. Throughout our experiments, we employed \textit{qwen-turbo\footnote{We call the qwen-turbo model from Alibaba Cloud via API.}} as the core LLM, generating expansions for both queries and documents. This model was selected for its fast response times, proven effectiveness, and low computational cost. 

\textbf{Base Retrieval Models}. For sparse retrieval, we used \textit{BM25} \cite{robertson2009probabilistic} as the foundational method. For dense retrieval, we used the \textit{multilingual-e5-base\footnote{https://huggingface.co/intfloat/multilingual-e5-base}} \cite{wang2024multilingual} model.

\textbf{Query Expansion and Document Expansion Baselines}
We compare our proposed TCDE with strong baselines in both query and document expansion. To ensure fair comparisons, we reproduce all methods using the experimental settings and configurations specified in their original papers.

Our query expansion baselines include four state-of-the-art methods: \textbf{Q2D} \cite{wang2023query2doc}, which generates a pseudo-document for expansion; \textbf{Q2C} \cite{jagerman2023query}, which employs chain-of-thought reformulation; \textbf{GRF} \cite{10.1145/3539618.3591992}, which produces a multifaceted expansion; and \textbf{CSQE} \cite{lei-etal-2024-corpus}, a retrieval-augmented method designed to mitigate hallucinations. Additionally, given TCDE's dual design, we include a leading document expansion method, \textbf{docT5query} \cite{nogueira2019doc2query}, using the official author-released checkpoint. As this model is trained on the MS MARCO corpus, we evaluated it only on corresponding benchmarks and excluded it from the BEIR suite to ensure a fair comparison.

\textbf{Implementation Details.} Our experimental pipeline begins by using \textit{qwen-turbo} to expand the documents for each dataset, setting the number of generated topics to 5 for each document. Then, for each query, we expand it using the same LLM, with the number of expanded topics also set to 5. Following this two-stage expansion, the expanded query is used to retrieve from the corpus of expanded documents. For our sparse retrieval experiments, we adopt the BM25s library \cite{lu2024bm25s}. For dense retrieval experiments and evaluation, we use the BEIR toolkit\footnote{\url{https://github.com/beir-cellar/beir}} \cite{thakur2021beir}.

\begin{table*}[!htbp]
\centering
\setlength{\tabcolsep}{1pt}
\begin{adjustbox}{width=0.9\textwidth}
\begin{tabularx}{\textwidth}{l|*{12}{Y}}
\toprule
\textbf{Methods} & \textbf{AA} & \textbf{CF} & \textbf{DB} & \textbf{Fe} & \textbf{FQ} & \textbf{HQ} & \textbf{NF} & \textbf{NQ} & \textbf{SD} & \textbf{SF} & \textbf{TC} & \textbf{WT} \\
\midrule
\multicolumn{13}{c}{\textbf{\textit{Sparse Retrieval}}} \\
\midrule
BM25        & 0.4874 & 0.1372 & 0.3045 & 0.5036      & 0.2532 & 0.5851 & 0.3180 & 0.2916 & 0.1565 & 0.6791 & 0.6099 & 0.3325 \\
\quad+Q2D        & 0.4260 & 0.2274 & 0.3199 & 0.6466      & 0.2404 & 0.5157 & 0.3367 & 0.4153 & 0.1530 & 0.6907 & 0.6350 & 0.3610 \\
\quad+Q2C        & 0.4367 & 0.2201 & 0.3533 & 0.6309      & 0.2558 & 0.5686 & 0.3336 & 0.4501 & 0.1554 & 0.7049 & 0.6848 & 0.3670 \\
\quad+GRF        & 0.4612 & \textbf{0.2523} & \textbf{0.3955} & \textbf{0.7433}      & 0.2769 & \textbf{0.6594} & \textbf{0.3740} & \textbf{0.5059} & \textbf{0.1719} & 0.7255 & 0.7480 & 0.4169 \\
\quad+CSQE       & 0.4950 & 0.2244 & 0.3840 & 0.6493      & 0.2943 & 0.6327 & 0.3620 & 0.4636 & 0.1692 & 0.7206 & \textbf{0.7674} & \textbf{0.4622} \\
\quad+TCDE       & \textbf{0.5135} & 0.2318 & 0.3798 & 0.7307      & \textbf{0.3165} & 0.6431 & 0.3581 & 0.4430 & 0.1651 & \textbf{0.7371} & 0.7189 & 0.3790 \\
\midrule
\multicolumn{13}{c}{\textbf{\textit{Dense Retrieval}}} \\
\midrule
E5-Base     & 0.4783 & 0.2196 & 0.3262 & 0.6832      & 0.3337 & 0.6715 & 0.3113 & 0.5588 & 0.1621 & 0.6887 & 0.3927 & 0.1285 \\
\quad+Q2D        & 0.4615 & \textbf{0.2667} & 0.3889 & 0.7338      & 0.3503 & 0.5765 & 0.3229 & 0.5343 & 0.1704 & 0.7069 & 0.6216 & 0.2279 \\
\quad+Q2C        & 0.3638 & 0.2449 & 0.3241 & 0.6865      & 0.2494 & 0.5036 & 0.2965 & 0.4864 & 0.1348 & 0.7023 & 0.5579 & 0.1771 \\
\quad+GRF        & 0.2974 & 0.2619 & 0.2900 & 0.6547      & 0.2547 & 0.3586 & 0.2758 & 0.3616 & 0.1380 & 0.6948 & 0.5863 & 0.2115 \\
\quad+CSQE       & 0.4123 & 0.2520 & 0.3990 & 0.7189      & 0.2898 & 0.6651 & 0.3439 & 0.5668 & 0.1594 & 0.7146 & 0.5532 & \textbf{0.2330} \\
\quad+TCDE       & \textbf{0.5081} & 0.2477 & \textbf{0.4184} & \textbf{0.7589}      & \textbf{0.3653} & \textbf{0.6773} & \textbf{0.3524} & \textbf{0.5820} & \textbf{0.1845} & \textbf{0.7325} & \textbf{0.6487} & 0.1885 \\
\bottomrule
\end{tabularx}
\end{adjustbox}
\caption{NDCG@10 results on the BEIR benchmark across 12 datasets. Bold indicates the best performance per dataset within each retrieval paradigm (sparse or dense). AA=ArguAna, CF=Climate-FEVER,  DB=DBPedia-Entity, Fe=FEVER, FQ=FiQA, HQ=HotpotQA, NF=NFCorpus, NQ=NaturalQuestions, SD=SCIDOCS, SF=SciFact, TC=TREC-COVID, WT=Webis-Touche2020.}
\label{tab:beir_results}
\end{table*}

\begin{table*}[!htbp]
    \centering
    \setlength{\tabcolsep}{4pt}
    \resizebox{0.9\textwidth}{!}{
    \begin{tabular}{l|ccc|ccc|ccc}
        \toprule
        \multirow{2}{*}{\textbf{Methods}} & \multicolumn{3}{c|}{\textbf{MS MARCO Dev}} & \multicolumn{3}{c|}{\textbf{TREC DL'19}} & \multicolumn{3}{c}{\textbf{TREC DL'20}} \\
         & N@10 & M@10 & R@1k & N@10 & M@10 & R@1k & N@10 & M@10 & R@1k \\
        \midrule
        \multicolumn{10}{c}{\textit{\textbf{Sparse Retrieval}}} \\
        \midrule
        BM25     & 0.2256 & 0.1762 & 0.8566 & 0.4734 & 0.1059 & 0.7322 & 0.4787 & 0.1362 & 0.7456 \\
        \quad+TQE     & 0.2319 & 0.1822 & 0.9056 & 0.6223 & 0.1440 & 0.8421 & 0.5904 & 0.1812 & 0.8531 \\
        \quad+TDE     & 0.2559 & \textbf{0.2016} & 0.8891 & 0.5344 & 0.1273 & 0.7767 & 0.4950 & 0.1435 & 0.7646 \\
        \quad+TCDE    & \textbf{0.2549} & 0.2001 & \textbf{0.9270} & \textbf{0.6657} & \textbf{0.1626} & \textbf{0.8637} & \textbf{0.6039} & \textbf{0.1829} & \textbf{0.8588} \\
        \midrule
        \multicolumn{10}{c}{\textit{\textbf{Dense Retrieval}}} \\
        \midrule
        E5-Base     & \textbf{0.4134} & \textbf{0.3439} & \textbf{0.9757} & 0.6943 & 0.1416 & 0.7735 & 0.6714 & 0.1974 & 0.7735 \\
        \quad+TQE     & 0.3942 & 0.3239 & 0.9755 & \textbf{0.7478} & \textbf{0.1736} & 0.8400 & 0.7131 & \textbf{0.2238} & 0.8188 \\
        \quad+TDE     & 0.4066 & 0.3360 & 0.9752 & 0.6874 & 0.1435 & 0.7936 & 0.6890 & 0.2071 & 0.7847 \\
        \quad+TCDE    & 0.3857 & 0.3155 & 0.9742 & 0.7235 & 0.1630 & \textbf{0.8554} & \textbf{0.7178} & 0.2222 & \textbf{0.8268} \\
        \bottomrule
    \end{tabular}
    }
    \caption{Ablation study of different topic-centric expansion components on sparse and dense retrieval baselines. We evaluate the impact of Topic-Centric Query Expansion (TQE), Topic-Centric Document Expansion (TDE), and Topic-Centric Dual Expansion (TCDE) on BM25 and E5-Base. Experiments are conducted on MS MARCO Dev, TREC DL'19, and TREC DL'20 datasets, with NDCG@10, MAP@10, and Recall@1k as evaluation metrics.}
    \label{tab:ablation_study}

\end{table*}

\subsection{Web Search Results}

As shown in \autoref{tab:results_msmarco}, our TCDE framework demonstrates strong performance gains across both sparse and dense retrieval paradigms, validating our core approach.

In the sparse retrieval setting, TCDE substantially outperforms the BM25 baseline and achieves state-of-the-art performance on the TREC DL'19 dataset. This confirms its ability to enrich traditional lexical models with effective semantic signals.

TCDE's advantages are most pronounced in the dense retrieval setting, where it establishes new state-of-the-art results across all metrics on both TREC DL'19 and DL'20. This dominant performance highlights the powerful synergy between our topic-centric expansion and dense embedding spaces. While all expansion methods show a slight performance drop on the highly-tuned MS MARCO Dev set, TCDE exhibits the most resilience, with the smallest degradation among all baselines. This indicates its topic-centric expansion is more robust and less noisy than other expansion strategies.

Collectively, these results validate that a topic-centric, dual expansion is a powerful strategy for enhancing modern retrieval systems.

\begin{figure*}[!htbp]
    \centering
    
    \begin{subfigure}{\textwidth}
        \centering
        \scalebox{0.95}{ 
            \begin{minipage}{0.3\textwidth}
                \includegraphics[width=\linewidth]{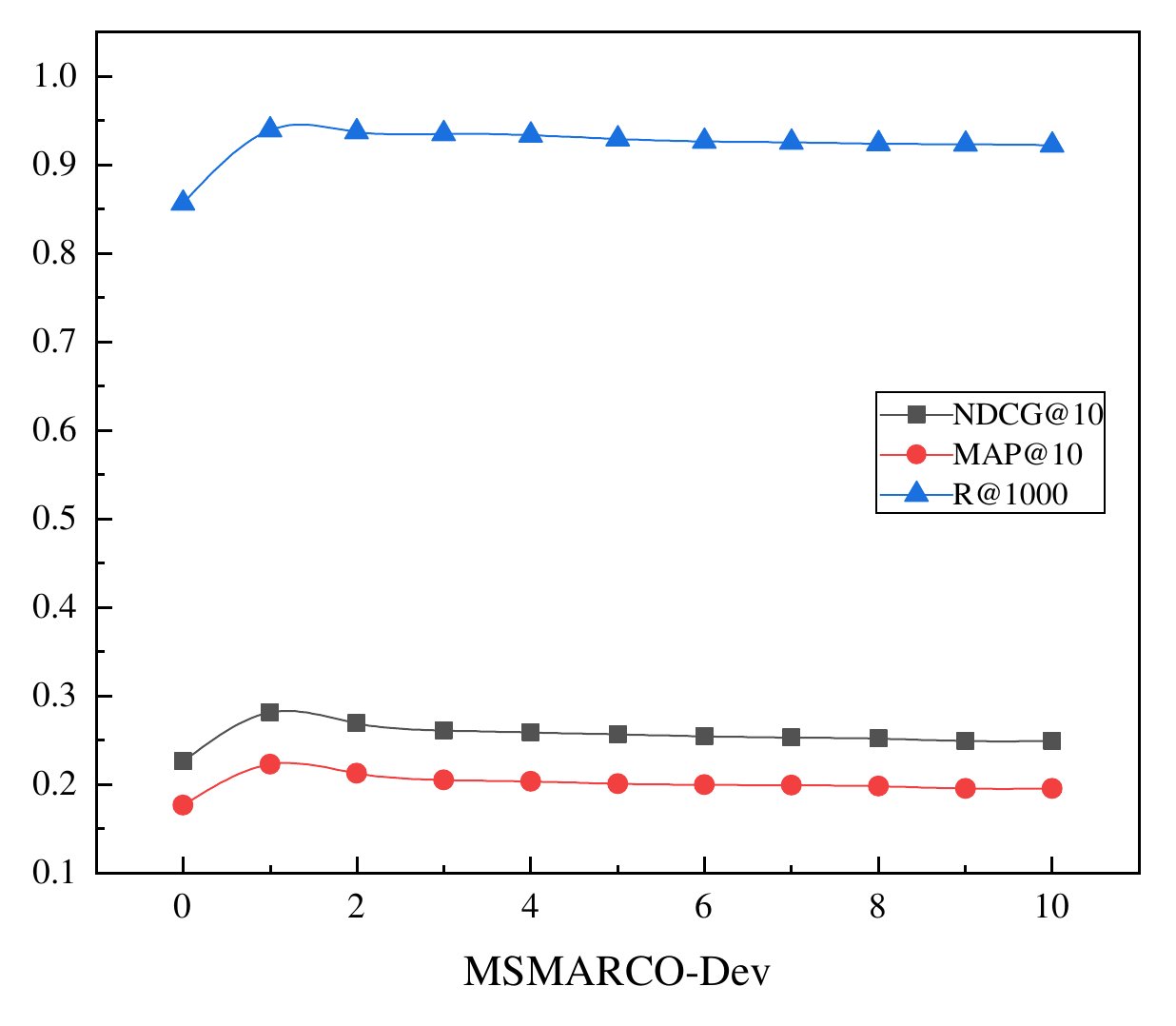}
            \end{minipage}\hfill
            \begin{minipage}{0.3\textwidth}
                \includegraphics[width=\linewidth]{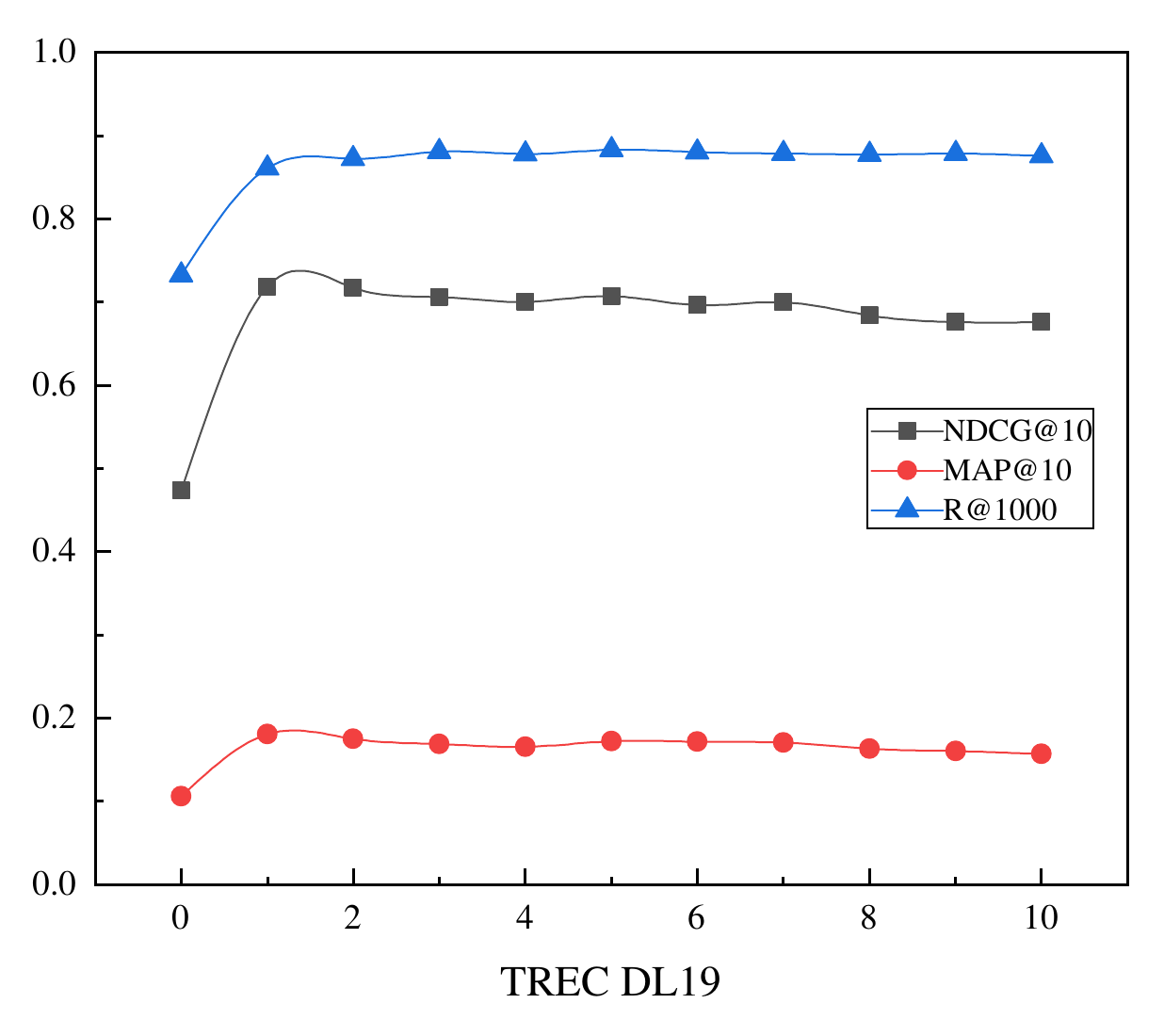}
            \end{minipage}\hfill
            \begin{minipage}{0.3\textwidth}
                \includegraphics[width=\linewidth]{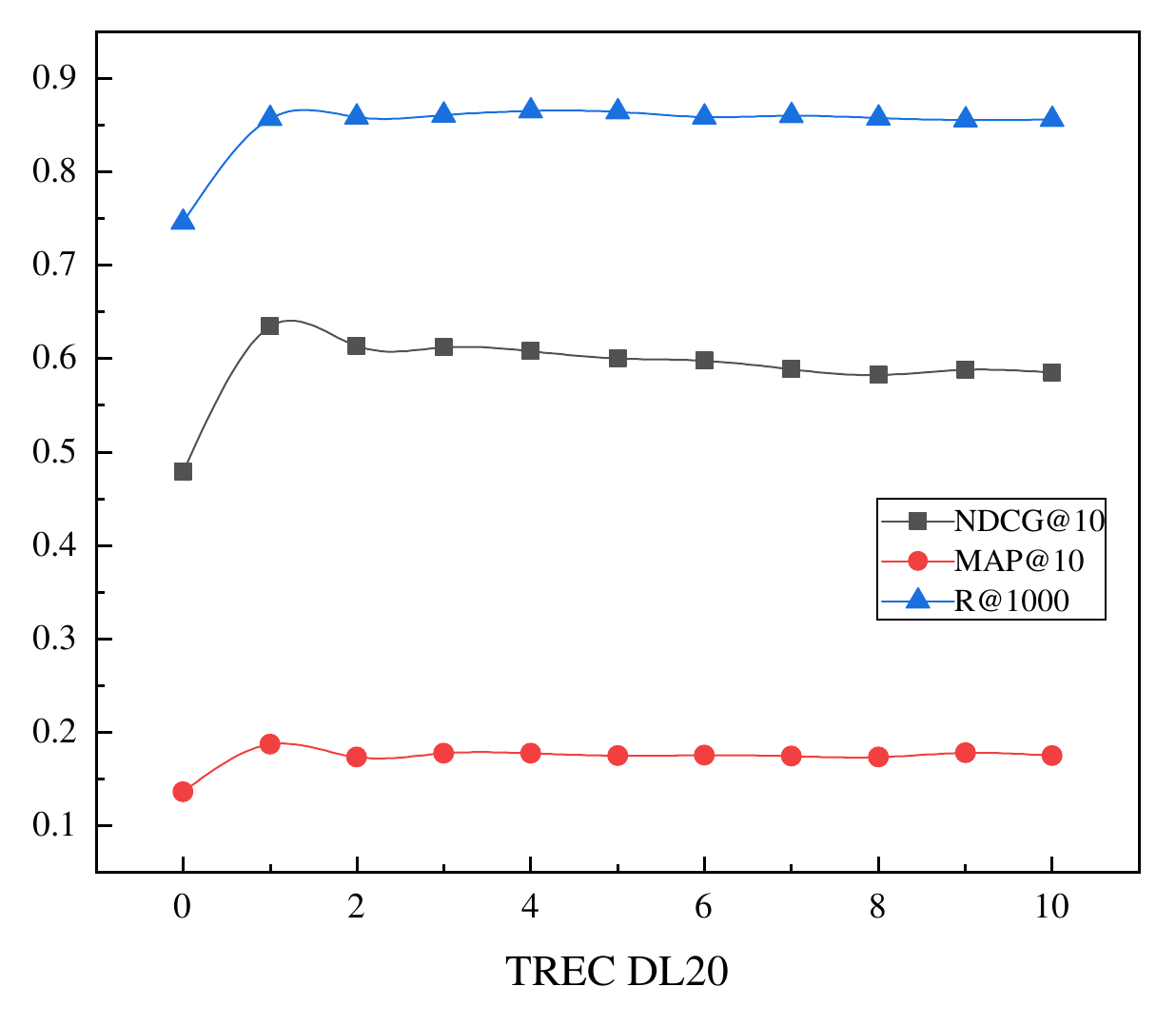}
            \end{minipage}
        }
        \caption{Impact of query expanded topics number} 
        \label{fig:row1_sub}
    \end{subfigure}

    \begin{subfigure}{\textwidth}
        \centering
        \scalebox{0.95}{ 
            \begin{minipage}{0.3\textwidth}
                \includegraphics[width=\linewidth]{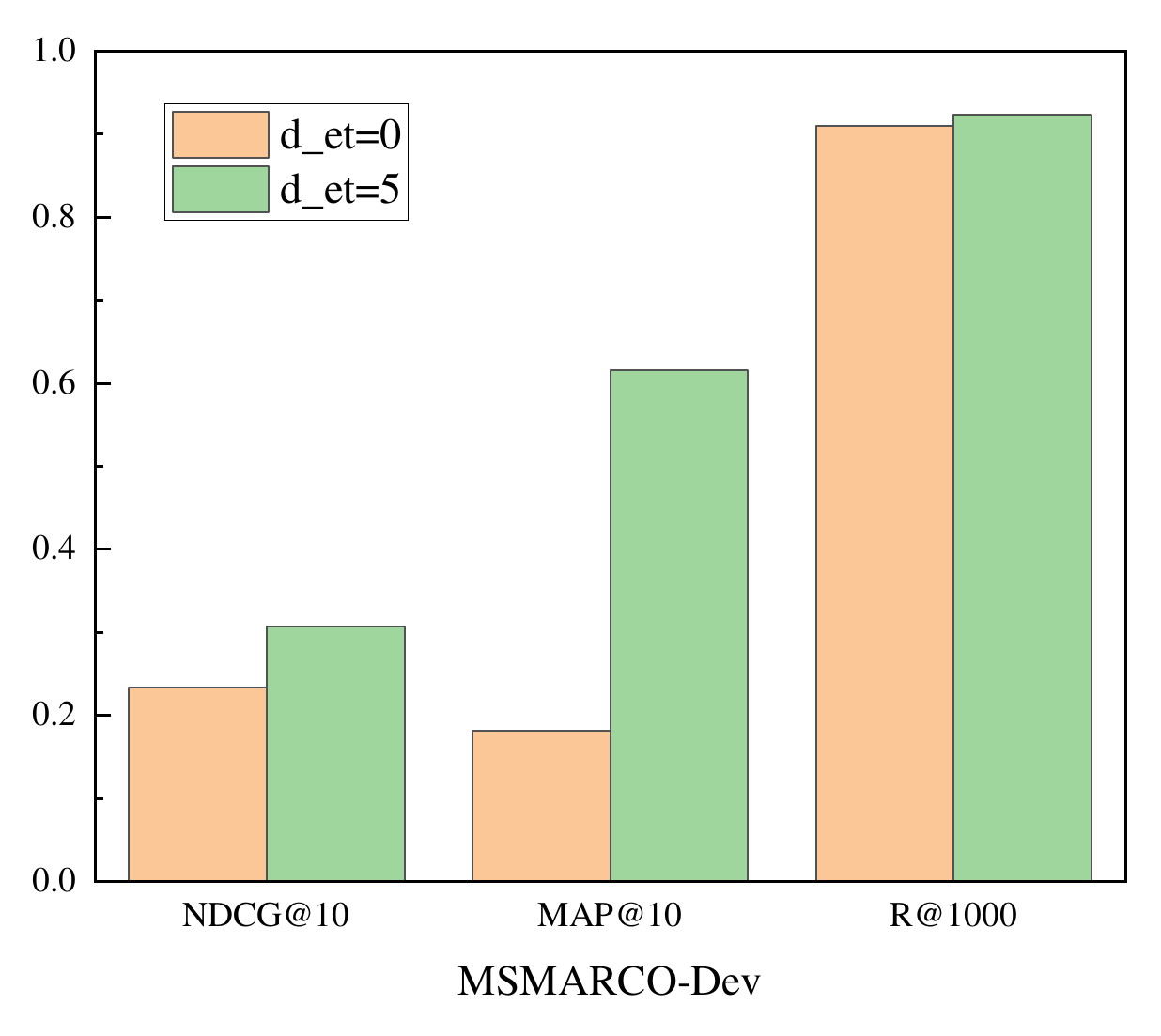}
            \end{minipage}\hfill
            \begin{minipage}{0.3\textwidth}
                \includegraphics[width=\linewidth]{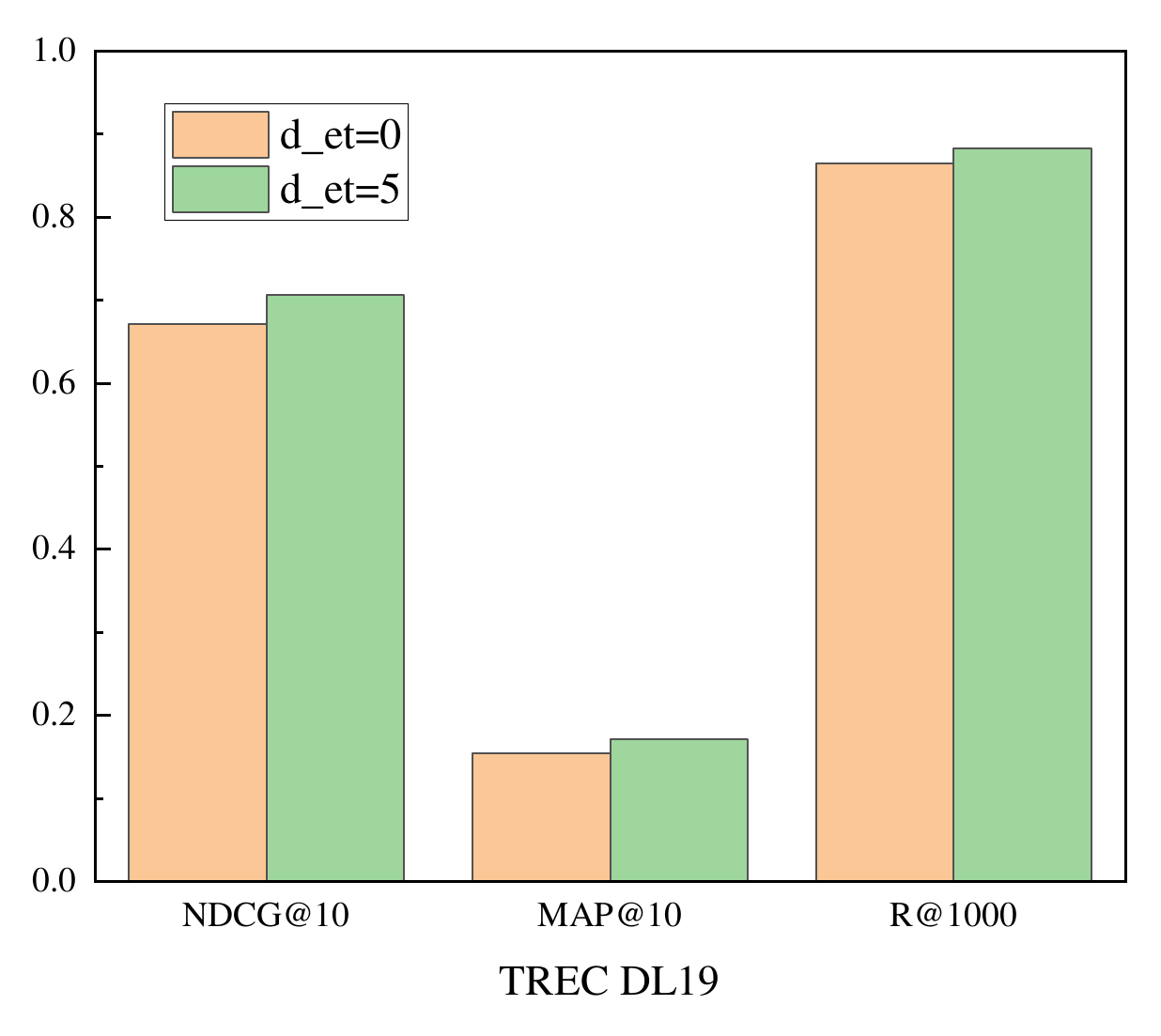}
            \end{minipage}\hfill
            \begin{minipage}{0.3\textwidth}
                \includegraphics[width=\linewidth]{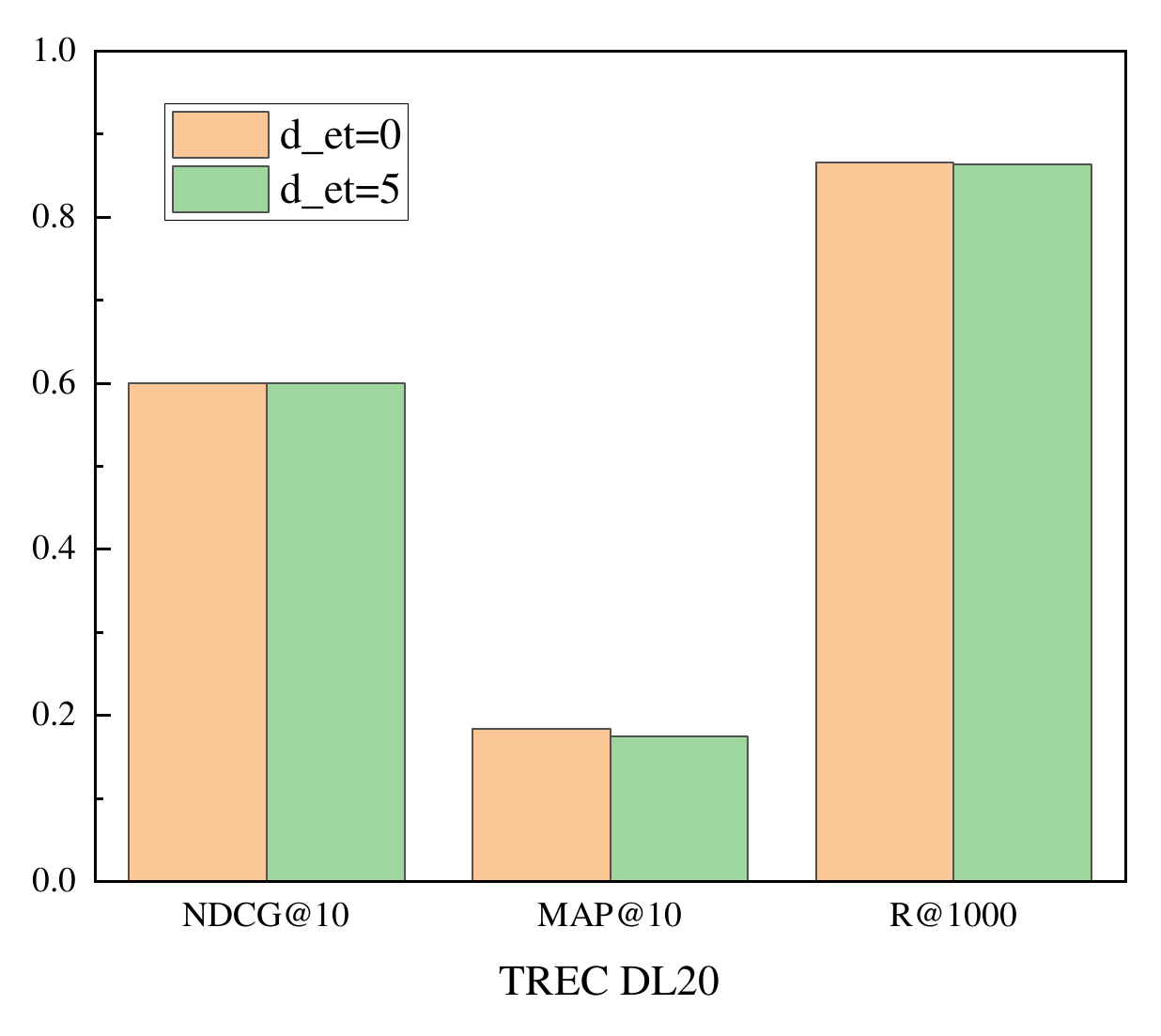}
            \end{minipage}
        }
        \caption{Impact of document expanded topics number} 
        \label{fig:row2_sub}
    \end{subfigure}

    \caption{Sensitivity analysis of the number of expansion topics. (a) The impact of varying the number of query expansion topics ($N_q$) from 0 to 10, with the number of document expansion topics ($N_d$) fixed at 5. (b) A comparison of performance with and without document expansion ($N_d$=0 vs. $N_d$=5), with the number of query expansion topics ($N_q$) fixed at 5. The analysis is conducted across all three datasets.} 
    \label{fig:impact_of_generated_topic_numbers}
\end{figure*}

\subsection{BEIR Retrieval Results}
To evaluate the generalization capabilities of our method, we conduct experiments on a diverse subset of 12 datasets selected from the BEIR benchmark. The results, presented in \autoref{tab:beir_results}, demonstrate that our proposed TCDE framework is a highly effective and versatile expansion strategy across both sparse and dense retrieval paradigms.

In the sparse retrieval setting, TCDE establishes itself as a top-tier performer. It achieves the highest NDCG@10 score on 3 of the 12 tested datasets (ArguAna, FiQA, and SciFact). While specialized methods like GRF and CSQE achieve peak performance on specific datasets, TCDE demonstrates superior robustness and generalization. In pairwise comparisons between TCDE and GRF, TCDE outperforms the strong GRF baseline on 8 of the 12 selected datasets, suggesting that its topic-centric expansion provides a more consistently beneficial signal across a wider variety of domains.

TCDE shows a dominant advantage in dense retrieval, which proves that our topic expansion strategy is highly effective at guiding the semantic matching process of dense embeddings. It achieves new state-of-the-art results on a remarkable 9 out of the 12 chosen datasets, decisively outperforming all other expansion methods. This strong performance extends across a wide range of tasks, from fact verification (FEVER, SciFact) to question answering (NaturalQuestions, HotpotQA). The substantial performance gains over the powerful E5-Base model, such as the +7.6\% absolute improvement on FEVER, strongly suggest that TCDE's explicit thematic signals effectively guide the dense retriever to discern true topic-centric alignment.

In summary, the comprehensive evaluation on this challenging subset of BEIR validates TCDE as a powerful and general-purpose expansion framework. Its consistent performance in the sparse setting, combined with its dominant results in the dense setting, confirms that our topic-centric, dual strategy is a significant step forward for modern retrieval systems.

\subsection{Ablation Study}
Our ablation study, presented in \autoref{tab:ablation_study}, dissects the individual contributions of our TCDE's components. The results indicate that both Topic-Centric Query Expansion (TQE) and Document Expansion (TDE), when applied in isolation, provide notable improvements over the base retrievers, with TQE often being the more dominant contributor, especially in the dense paradigm.

However, the most crucial finding is that the full TCDE model, which synergistically combines both components, consistently outperforms either unidirectional expansion strategy across nearly all datasets and metrics. This demonstrates a clear synergistic effect, confirming that the dual architecture is critical to TCDE's overall effectiveness and superior performance.

\subsection{Impact of generated topic numbers}
To gain deeper insights into our proposed method, we conducted a parameter sensitivity analysis focused on the number of topics used for expansion. Our investigation was twofold. First, we examined the impact of the number of query topics by varying it within the range of [0, 10], while holding the number of document topics constant at 5. Second, we assessed the impact of the number of document topics. As document expansion is computationally intensive, we limited this comparison to two settings (0 and 5 topics), while fixing the number of query topics at 5. The results of this analysis are illustrated in \autoref{fig:impact_of_generated_topic_numbers}.

\autoref{fig:row1_sub} reveals that performance dramatically improves when using a small number of query topics (e.g., $N_q$=2) and then plateaus, indicating that a few topics are sufficient to capture the query's intent. Adding more topics provides diminishing returns. \autoref{fig:row2_sub} confirms that document expansion is also beneficial, as the configuration with 5 document topics ($N_d$= 5) consistently outperforms no document expansion ($N_d$=0) across all metrics. In summary, the analysis shows that both expansion types are effective and that our model is robust to these hyperparameters. A setting of $N_q=5$ and $N_d=5$ is a reliable choice to achieve strong performance.

\section{Conclusion}

In this paper, we introduce TCDE, a topic-centric dual expansion framework designed to strengthen the topic-centric alignment between queries and documents in information retrieval. Our evaluation reveals two primary contributions. First, we establish that TCDE's focused, topic-centric expansion strategy is significantly more effective than traditional pseudo-relevance feedback, even when such feedback is generated by state-of-the-art Large Language Models. This underscores the core importance of topic-centric alignment over raw generative capacity for information retrieval. Second, we demonstrate that the framework's dual architecture is essential, consistently yielding substantial gains over any unidirectional approach. In conclusion, TCDE's effectiveness is not attributed to its underlying LLM alone, but to the critical synergy of integrating these two strategies. By guiding the LLM's power with a structured, dual, and topic-focused approach, TCDE offers a more robust and principled path for enhancing both queries and documents than methods that expand only one side.

\section*{Ethical Considerations}

\noindent\textbf{Scope and datasets.}
This work studies topic-centric dual expansion for information retrieval using publicly available benchmarks (e.g., MS MARCO, BEIR). We did not collect new user data, recruit human subjects, or access private logs. All experiments operate on public text corpora and their official queries/labels. No personally identifiable information (PII) was newly gathered or processed beyond what is already present in the benchmarks, and our preprocessing removes obvious PII when encountered.

\medskip\noindent\textbf{Privacy, licensing, and intellectual property.}
We comply with the licenses and terms of use of all datasets and third-party resources. Our expansions are generated via large language model (LLM) inference and are stored only for experimental evaluation. We do not redistribute original corpora beyond their licenses. Where applicable, we filter or truncate generated text to avoid copying long spans from proprietary sources and to minimize the risk of re-exposing training data. No end-user identifiers were used, and all API keys or credentials are kept outside the artifact.

\medskip\noindent\textbf{Bias, fairness, and topical drift.}
LLMs may amplify societal, topical, or domain biases and can introduce hallucinations or off-topic content. To mitigate these risks, our method constrains generation to topic-focused summaries/sentences, and rejects empty or clearly off-topic expansions. We report failures and ablations on topic drift (e.g., effects of the number/length of expansions) and encourage downstream deployers to layer additional safety filters (toxicity, hate/harassment, and unsafe instruction detectors) appropriate to their application domain.

\medskip\noindent\textbf{Reproducibility and transparency.}
To support reproducibility while preserving double-blind review, we provide detailed prompts, hyperparameters, and evaluation scripts will release them upon acceptance. No identifying URLs, organization names, or commit histories are included in the submission materials. After publication, we intend to release artifacts consistent with dataset licenses and community norms, and we will honor takedown requests for inadvertent inclusion of sensitive material.

\medskip\noindent\textbf{Limitations.}
Our expansions are textual heuristics that may not capture all facets of relevance; they can still drift or under-represent minority viewpoints in specialized domains. We therefore recommend domain-expert review for safety-critical deployments and encourage future work on bias auditing, multilingual fairness, and controllable generation constraints.

\clearpage
\bibliographystyle{ACM-Reference-Format}
\bibliography{references} 




\end{document}